\documentclass[useAMS,usenatbib]{mn2e}
\usepackage{times}
\usepackage{epsfig}
\title[A hybrid model for quasar optical variability]{An Accretion Disc-Irradiation Hybrid Model for The Optical/UV Variability in Radio-Quiet Quasars}
\author[]{Hui Liu$^{1,2}$, Shuang-Liang Li$^{1,}$\thanks{Corresponding author: lisl@shao.ac.cn}, Minfeng Gu$^1$\\
$^1$ Key Laboratory for Research in Galaxies and Cosmology, Shanghai Astronomical Observatory, Chinese Academy of Sciences,\\
80 Nandan Road, 200030 Shanghai, China; liuhui@shao.ac.cn; gumf@shao.ac.cn\\
$^2$ University of Chinese Academy of Science, 19A Yuquanlu, Beijing 100049, China;}

  \label{firstpage}
\begin{document}


\pagerange{\pageref{firstpage}--\pageref{lastpage}} \pubyear{2015}

\maketitle

\label{firstpage}

\begin{abstract}
The optical/ultraviolet (UV) variability of quasars has been discovered to be correlated with other quasar properties, such as luminosity, black hole mass and rest-frame wavelength. However, the origin of variability has been a puzzle so far. In this work, we upgrade the accretion disc model (Li \& Cao 2008), which assumed the variability is caused by the change of global mass accretion rate, by constraining the disc size to match the viscous timescale of accretion disc to the variability timescale observed and by including the irradiation/X-ray reprocessing to make the emitted spectrum become steeper. We find this hybrid model can reproduce the observed bluer-when-brighter trend quite well, which is used to validate the theoretical model by several works recently. The traditional correlation between the variability amplitude and rest-frame wavelength can also be well fitted by our model. In addition, a weak positive correlation between variability amplitude and black hole mass is present, qualitatively consistent with recent observations.

\end{abstract}


\section{Introduction\label{intro}}

Flux variability at all the bands is a well-known characteristic of active galactic nuclei (AGNs) \citep{Ulrich1997}. The optical/UV emission of quasars, varying at timescales of hours to decades, is believed to come from an optically thick and geometrically thin accretion disc. Therefore, the study on quasar variability can help to understand the accretion process therein.

The correlation of quasar variability amplitude and other properties has been investigated for several decades. An negative correlation between variability and luminosity was presented by numerous authors adopting various samples \citep{Cid1996,Cristiani1996,Vanden2004,Wilhite2008,Ai2010,Zuo2012,Guo2014}. By matching quasars from the Quasar Equatorial Survey Team Phase 1 (QUEST1) variability survey with broad-line objects from the Sloan Digital Sky Survey, \citet{Wold2007} reported a positive correlation between variability and black hole mass, which was confirmed subsequently by larger samples \citep{Wilhite2008,Bauer2009}. Except for luminosity and black hole mass, variability is also found to be correlated with redshift and wavelength \citep*[e.g.,][]{Vanden2004,Zuo2012}. Another prominent feature of quasar variability is whose spectral usually tends to be bluer when brighter \citep*[e.g.,][]{Vanden2004,Sakata2011,Schmidt2012,Zuo2012,Guo2014,Kokubo2014,Ruan2014,Sun2014,Cai2016}, though a small fraction of quasars show a redder-when-brighter trend \citep{Schmidt2012,Kokubo2014}.

The origin of quasar variability has remained unclear so far. A number of models have been produced to explore the physical process therein, such as, accretion disc with variable mass accretion rate \citep{Pereyra2006,Lisl2008,Gu2013}, inhomogeneous accretion disc \citep{Kawaguchi1998,Dexter2011}, X-ray reprocessing/irradiation \citep{Tomita2006,Cackett2007,Gil-Merino2012,Chelouche2013}, and gravitational microlensing effect \citep{Hawkins1993,Hawkins2002}. Recently, the bluer-when-brighter trend has been used to validate quasar variability model. Several works \citep{Schmidt2012,Kokubo2014,Ruan2014} argued that this bluer-when brighter trend observed can't fully be reproduced by the change of global mass accretion rate because, for example, the accretion disc model can't reproduce the observed steep `relative variability spectrum' after correcting the extinction from both the host-galaxy and our Galaxy \citep{Ruan2014}. Instead, they suggested the inhomogeneous accretion disc model produced by \citet{Dexter2011} can work well. Nevertheless, \citet{Kokubo2015} argued that this model seems to be ruled out because it can't explain the tight inter-band correlation in quasar variability.

Accretion disc model was found to be able to explain the correlations between variability amplitude and other quasar properties quite well \citep{Pereyra2006,Lisl2008,Sakata2011,Zuo2012}. However, there is still two points required for further attention. At first, the viscous timescale of an accretion disc ($\sim 10^3$ years at $R=10^3 R_{\rm g}$ for a $10^8 M_\odot$ black hole, where $R_{\rm g}$ is the Schwarzchild radius) is much longer than the quasar variability timescale observed (days $\sim$ decades). Secondly, as mentioned above, the accretion disc model can't fit the observed difference spectrum well. On the other hand, a correlation between optical and X-ray variability is observed in quasar variability \citep*[e.g.,][]{Breedt2009,Cameron2012}, which can be explained by the X-ray reprocessing model \citep{Tomita2006,Cackett2007,Gil-Merino2012,Chelouche2013}.

Therefore, an accretion disc-irradiation hybrid model is adopted in this work to improve the accretion disc model of \citet{Lisl2008} at two points: 1) we constrain the accretion disc size to make sure that the viscous timescale at outer radius of disc is consistent with the observed variability timescale; 2) we consider the effect of irradiation from an X-ray point source above the inner disc, which can help to increase the flux variation in UV bands and make the spectrum become bluer.

\section{Accretion disc-irradiation hybrid model}\label{models}

The energy flux of a geometrically thin and optically thick standard thin disc is given by:
\begin{equation}
F_{\rm disc}={\sigma}T_{\rm eff}^{4}=\frac{3GM\dot{M}}{8{\pi}R^{3}}\left(1-\sqrt{\frac{R_{\rm in}}{R}}\right),
\label{energyeq}
\end{equation}
where $T_{\rm eff}$ is the effective temperature of disc, $M$ is the black hole mass, $\dot{M}$ is the mass accretion rate and $R_{\rm in}=3 R_{\rm g}$ is the inner radius of disc \citep{Kato1998}.

An X-ray point source above the accretion disc is usually adopted to describe the irradiation/X-ray reprocessing though its location and size is still a puzzle \citep*[e.g.,][]{Wilkins2012}. In this work, we simply adopt an X-ray point source located on the rotational axis above the disc \citep{Blaes2004,Liu2011}. Thus, the irradiation flux to the surface of accretion disc can be expressed as (Fig. \ref{shiyitu}):
\begin{figure}
  \centering
  \includegraphics[width=8cm]{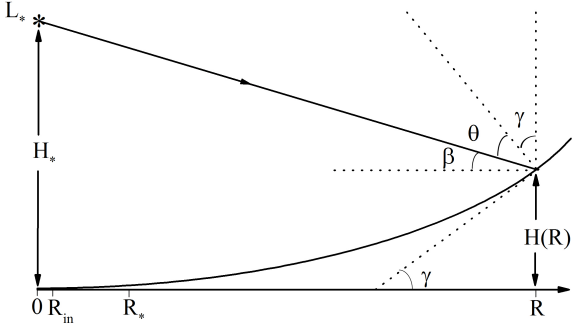}\\
 \caption{Schematic diagram for the accretion disc-irradiation hybrid model. The X-ray point source locates on the rational axis above the disc.}
 \label{shiyitu}
\end{figure}
\begin{equation}
F_{\rm irr}=\frac{L_{*}(1-a)\cos{\theta}}{4{\pi}R^{2}},
\end{equation}
where $L_{*}$ is the luminosity of X-ray point source and $a$ is the albedo of accretion disc. ${\theta}={\pi}/{2}-({\beta}+\gamma)$ is the angle between normal of disc and the light ray, where $\tan{\beta}={|H_{*}-H|}/{R}$ and $\tan{\gamma}={dH}/{dR}\sim{g{{H}/{R}}}$, here $g\sim 9/8$ and $\sim9/7$ for $R\ll H_{\ast}$ and $R\gg H_{\ast}$, respectively \citep{Kato1998}. We adopt $g=1.2$ in this work for simplicity. Thus $\cos{\theta}$ can be given by:
\begin{equation}
 \cos{\theta}=\frac{R|H_{*}-H|+gH}{\left(R^{2}-gH|H_{*}-H|\right)\sqrt{1+\left(\frac{R|H_{*}-H|+gH}{R^{2}-gH|H_{*}-H|}\right)^{2}}}.
 \label{cos}
\end{equation}

In order to solve the long viscous timescale problem at large radii, we shrink the outer radius of accretion disc from $R_{\rm out}$ to $R_{\ast}$ (Fig. \ref{shiyitu}), at which the viscous timescale is limited to $10$ years (the typical quasar variability timescale observed). For an initial mass accretion rate $\dot{M}_{1}$, the total energy flux radiated from an accretion disc and X-ray point source is
\begin{eqnarray}
F_{1}&=&{\sigma}T_{\rm eff,1}^{4}=F_{\rm disc,1}+F_{\rm irr,1}\nonumber\\
&=&\frac{3GM\dot{M}_1}{8{\pi}R^{3}}\left(1-\sqrt{\frac{R_{\rm in}}{R}}\right)+\frac{L_{*,1}(1-a)\cos{\theta}}{4{\pi}R^{2}}.
\label{energy1}
\end{eqnarray}

Assuming that the global mass accretion rate varies from $\dot{M}_{1}$ to $\dot{M}_{2}$ and the luminosity of X-ray point source varies from $L_{*,1}$ to $L_{*,2}$, the energy flux can be calculated with
\begin{eqnarray}
F_{2}&=&{\sigma}T_{\rm eff,2}^{4}=F_{\rm disc,2}+F_{\rm irr,2}\nonumber\\
&=&\frac{3GM\dot{M}_2}{8{\pi}R^{3}}\left(1-\sqrt{\frac{R_{\rm in}}{R}}\right)+\frac{L_{*,2}(1-a)\cos{\theta}}{4{\pi}R^{2}}
\label{energy2}
\end{eqnarray}
when $R_{\rm in}<R<R_{\ast}$. For $R_{\ast}<R<R_{\rm out}$, however, we maintain the mass accretion rate onto $\dot{M}_{1}$ due to its large viscous timescale. Thus the the energy flux is given by
\begin{eqnarray}
F_{2}&=&{\sigma}T_{\rm eff,2}^{4}=F_{\rm disc,1}+F_{\rm irr,2}\nonumber\\
&=&\frac{3GM\dot{M}_1}{8{\pi}R^{3}}\left(1-\sqrt{\frac{R_{\rm in}}{R}}\right)+\frac{L_{*,2}(1-a)\cos{\theta}}{4{\pi}R^{2}}.
\label{energy3}
\end{eqnarray}
With equations (\ref{energy1}),(\ref{energy2}), (\ref{energy3}), the spectrum of the accretion disc can be calculated with
\begin{equation}
f_{\nu}=\frac{4{\pi}\cos i\nu^{3}}{c^{2}D^{2}}\int_{R_{\rm in}}^{R_{\rm out}}\frac{RdR}{e^{h{\nu}/{kT_{\rm eff}}}-1},
\label{spectrum}
\end{equation}
where $i$ is the inclination of axis of the disc with respect to the line of sight, and $D$ is the distance from observer to black hole, $R_{\rm out}=1000R_{g}$ is the outer radius of disc and $h$ is the Plank's constant \citep{Frank2002}.

\section{results}\label{results}

\begin{figure*}
  \centering
  \includegraphics[width=16cm]{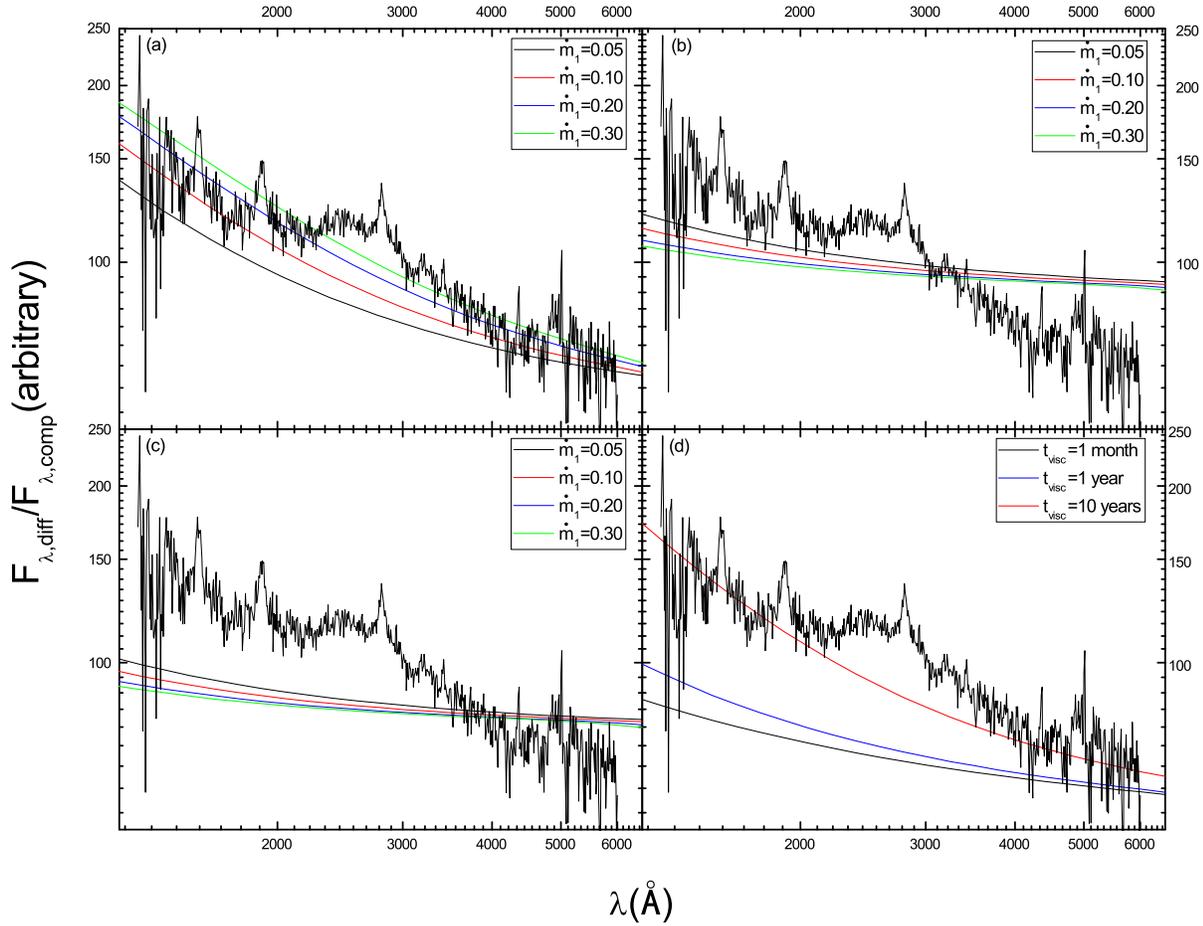}\\
  \caption{The relative variability spectrum as functions of rest-frame wavelength for different initial mass accretion rate $\dot{m}_{1}$ and different viscous timescale. The black curve represents the observed composite relative variability spectrum given in Fig.6 of \citet{Ruan2014}. Panel (a) represents our new model in this work, where the parameters $\alpha=0.1$, $a=0.4$, $L_{\ast}=0.3 L_{\rm bol}$ ($L_{\rm bol}=GM\dot{M}/2R_{\rm in}$), and $\delta{\dot{m}}= 100\%\dot{m}_{1}$ are adopted. Here   $\dot{m}={\dot{M}}/{\dot{M}_{\rm Edd}}, \dot{M}_{\rm Edd}=1.5\times{10^{18}}mgs^{-1}, m=M/M_{\odot}$. Panel (b) represents the standard accretion disc model (no limiting the disc size), where all the parameters are the same as panel (a) except that $\delta{\dot{m}}= 30\%\dot{m}_{1}$. Panel (c) represents the standard accretion disc model plus irradiation, where all the parameters are the same as panel (b). Panel (d) also represent our model but for different viscous timescale, where all the parameters are the same as panel (a) except that $\dot{m}=0.1$.}  \label{ruan}
\end{figure*}

\begin{figure*}
  \centering
  \includegraphics[width=16cm]{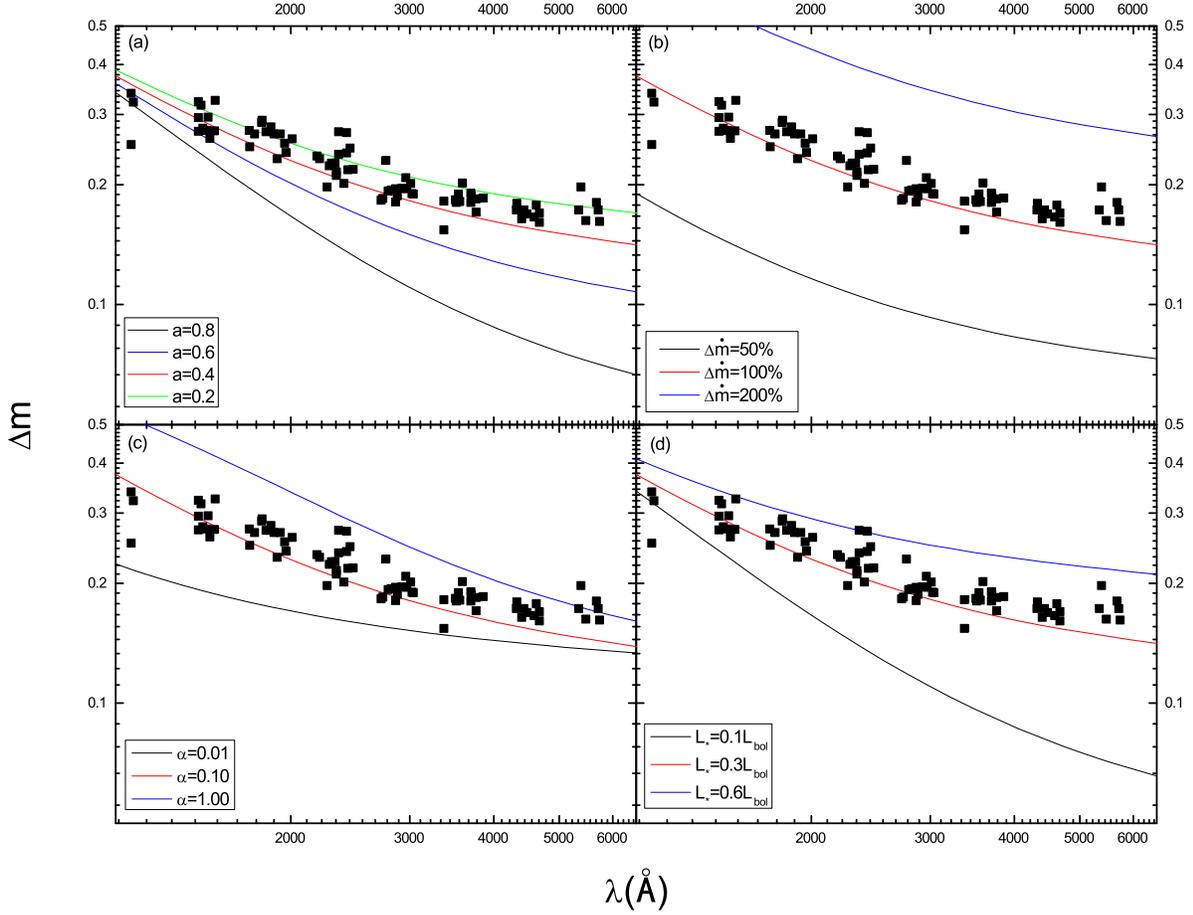}\\
  \caption{The variability amplitude as functions of rest-frame wavelength for different parameters, where $\delta \dot{m}=100\%\dot{m}_{1}$ except for panel (b). The black squares are the statistical results given in Fig.13 of \citet{Vanden2004}.}\label{vanden}
\end{figure*}

\begin{figure}
  \centering
  \includegraphics[width=8cm,height=6cm]{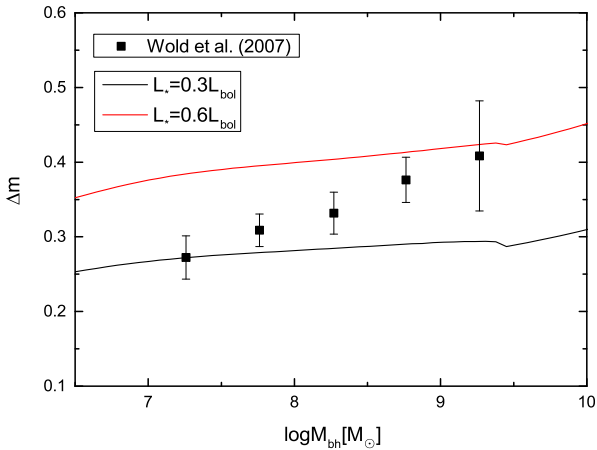}\\
  \caption{The variability amplitude in R-band as functions of black hole mass for different $L_{\ast}$, where all the parameters adopted are the same as Fig. \ref{ruan} except that $\dot{m}_{1}=0.1$ and $\delta\dot{m} = 2\dot{m}_{1}$. The five black squares with error bars are the statistical results given in Fig.5 of \citet{Wold2007} (see their paper for details.). The red and black lines are for $L_{\ast}=0.3L_{\rm bol}$ and $0.6L_{\rm bol}$, respectively.}\label{logmva}
\end{figure}
\begin{figure}
  \centering
  \includegraphics[width=8cm,height=6cm]{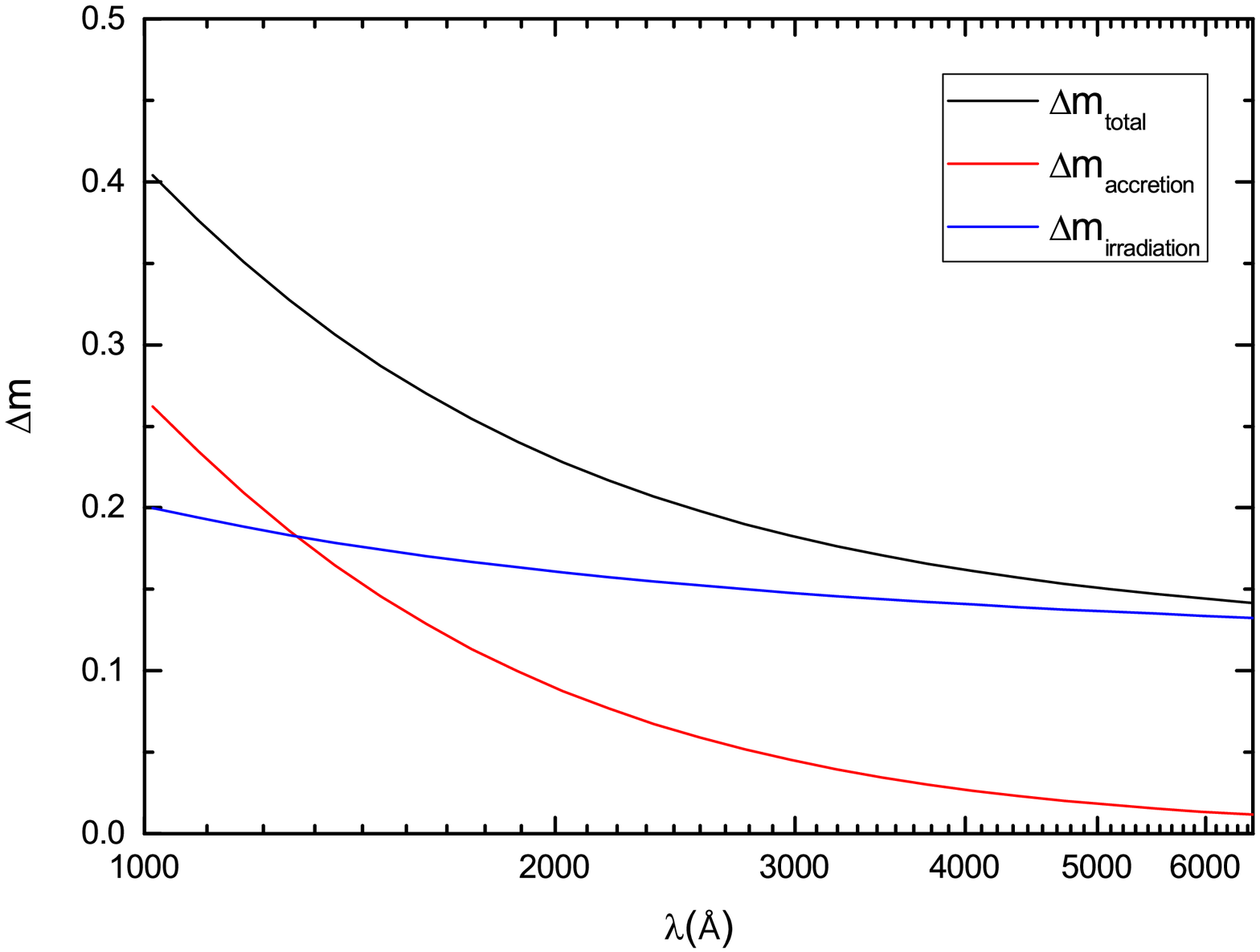}\\
  \caption{The variability amplitude as functions of rest-frame wavelength for the well-fitting parameters. The red and blue lines correspond to the variability from accretion disc and irradiation, respectively.}\label{parts}
\end{figure}

Given the black hole mass $M$, the mass accretion rate $\dot{M}$, the Shakura-Sunyaev parameter $\alpha$, the disc albedo $a$, and the luminosity of the X-ray point source $L_{*}$, we can solve the disc equations to get the scaled-height $H$ of the disc. Therefore, the disc spectrum can be calculated with equations (\ref{cos}) - (\ref{spectrum}). In all the calculations, the variability is caused by the change of mass accretion rate, and $H_{\ast}=10 R_{\rm g}$ and $M=10^8 M_{\odot}$ are always adopted.

The bluer-when-brighter trend observed was adopted to validate the theoretical model for quasar variability by several authors recently \citep{Schmidt2012,Ruan2014}. Therefore, we firstly compare our result with the `relative variability spectrum' of  \citet{Ruan2014} in Fig. \ref{ruan}a, where the relative variability spectrum means the spectral variability relative to the underlying spectra of quasars \citep*[the raito of the geometric mean composite difference spectrum to the geometric mean composite spectrum, see][for details]{Ruan2014}. It is found that the observed spectrum can be well reproduced by the results of our model (excluding the bump due to Fe II lines around Mg II 2800$\AA$), where the red line represents the well-fitting parameters. The standard accretion disc model can produce a bluer when brighter trend too since the disc temperature will increase with increasing mass accretion rate, which makes the spectrum become bluer. Therefore, we compare the results of standard accretion disc model with the observed relative variability spectrum in Fig. \ref{ruan}b. It is found that the former is much flatter than the latter, which is consistent with the results in several recent works \citep{Schmidt2012,Kokubo2014,Ruan2014}. Except for the standard accretion disc model adopted in Fig. \ref{ruan}b, we further explore the effects of limiting disc size in Fig. \ref{ruan}c by incorporating irradiation with the standard accretion disc model. The results are quite similar with Fig. \ref{ruan}b because the variability is dominated by the change of accretion disc. Therefore, only the hybrid model adopted in this work can fit the observed relative variability spectrum well. In most of our calculations, we adopt a viscous timescale of 10 years for the reason that the typical timescale of quasar variability is about several years. However, quasar also shows variability on timescales of months, weeks and sometime days. Thus we investigate the effects of viscous timescale on our results in Fig. \ref{ruan}d. Our model can be applied to quasars with all kinds of timescales, but the variability tends to be dominated by irradiation when the timescale is small. Therefore, we find the spectra become more redder when the viscous timescale is shorter. Limiting the disc size is somewhat similar with the fluctuation models \citep[e.g.,][]{Lyubarskii1997}, where the mass accretion rate is assumed to fluctuate and propagate on viscous timescales at all the radii. Thus the short variability will come from the rapid fluctuations at inner disc region.

In order to further validate our model, we compare the results of our model with the observed variability amplitude given by \citet{Vanden2004} in Fig. \ref{vanden}. It is found that our model can fit the observed variability quite well either, where the red lines are the well-fitting results. The same as Fig. \ref{ruan}, all the parameters corresponding to the well-fitting lines are conventional, i.e., $\alpha=0.1$, $a=0.4$, $\dot{m}=0.1$, and $L_{*}=0.3 L_{\rm Edd}$. Fig. \ref{vanden}a, \ref{vanden}b, \ref{vanden}c, and \ref{vanden}d are designed to explore the effects of different parameters, $a$, $\delta \dot{m}$, $\alpha$, and $L_{*}$, respectively. $\delta \dot{m}$, and $L_{*}$ are found to have more effects on the results than other parameters.

At last, we compare the positive correlation between variability amplitude and black hole mass \citep{Wold2007,Wilhite2008} with our model (Fig. \ref{logmva}). A somewhat weak positive correlation is present in our model comparing with the  observed correlation, where the dip corresponds to the black hole mass for which $R_{*}$ will be smaller than $R_{\rm in}$. In this case, the variability will totally come from irradiation. One possibility to increase this correlation is assuming that the accretion rate at inner radius can vary more than that at outer radius, as suggested by \citet{Wold2008}. Therefore, the disc region emitting at R band will move to smaller radii for larger mass black holes, resulting on the increase of variability at R band.

\section{conclusions and discussion}\label{conclusions and discussion}

In this work, we present an accretion disc-irradiation hybrid model for the optical/UV variability in quasars. In order to solve both the long viscous timescale trouble at large radius and the failure to fit the bluer-when brighter trend in UV bands, we upgrade the accretion disc model by constraining the disc size to reduce the viscous timescale and by including the irradiation/X-ray reprocessing from an X-ray point source above the disc to make the spectrum become steeper. It is found that our model can reproduce the `relative variability spectrum' of \citet{Ruan2014} quite well (Fig. \ref{ruan}a). The traditional positive correlation between variability amplitude and the rest-frame wavelength can be well fitted too (Fig. \ref{vanden}). The variation of mass accretion rate $\delta\dot{m}$ and the luminosity of X-ray point source $L_{\ast}$ are found to have more effects on our results than other parameters.

We include both the effects of accretion disc and X-ray point source in this work. The variability at EUV bands is found to be dominated by the contribution from accretion disc and the irradiation controls the variability when wavelength is smaller than 1300 ${\AA}$ (Fig. \ref{parts}). Thus, at R-band, the variability mainly comes from irradiation. Furthermore, the geometry of disc (e.g., flaring, humps or tapering of the disc) will also play an important role on the contribution of irradiation \citep{Lira2011}. But in this work, instead of assuming a disc shape in advance, the disc height is gotten by solving the disc equation \citep{Kato1998} plus irradiation. In addition, the contribution from accretion disc will disappear when the black hole mass $M \geq 10^{9}M_{\odot}$ due to its large viscous timescale. The observed timescale of quasar variability adopted is 10 years in all the calculations. Therefore, the effect of accretion disc will decrease if the observed timescale is smaller, and vice versa.

The hard X-ray bolometric correction $k_{\rm bol}=L_{\rm bol}/L_{\rm 2-10kev}$ is about 20 for X-ray selected type 1 AGN with Eddington ratio $\lambda_{\rm Edd}\sim 0.1$ \citep{Lusso2010}. Thus, the ratio of total X-ray to bolometric luminosity, which should be several times larger than $1/k_{\rm bol}$ $(=0.05)$, is roughly consistent with the luminosity of X-ray point source ($L_{\ast}=0.3 L_{\rm bol}$) adopted in most of our calculations. This value is also in compliance with that constrained from observations on X-ray reprocessing \citep*[e.g.,][]{Chelouche2013}.

 \section*{Acknowledgement}
We thank the reviewer for his/her very helpful report. J. Ruan is grateful for kindly providing their data on relative variability spectrum. H. Liu thanks H. Guo for helpful discussion. This work is supported by the NSFC (grants 11233006, 11373056, 11473054, and U1531245) and the Science and Technology Commission of Shanghai Municipality (13ZR1447000 and 14ZR1447100).

\label{lastpage}
\end{document}